# Forms of Quantum Nonseparability and Related Philosophical Consequences

## Vassilios Karakostas

Department of Philosophy and History of Science, University of Athens, Athens 157 71, Greece (E-mail: **vkarakos@cc.uoa.gr**).

**Abstract:** Standard quantum mechanics unquestionably violates the separability principle that classical physics (be it point-like analytic, statistical, or field-theoretic) accustomed us to consider as valid. In this paper, quantum nonseparability is viewed as a consequence of the Hilbert-space quantum mechanical formalism, avoiding thus any direct recourse to the ramifications of Kochen-Specker's argument or Bell's inequality. Depending on the mode of assignment of states to physical systems (unit state vectors versus non-idempotent density operators) we distinguish between strong/relational and weak/deconstructional forms of quantum nonseparability. The origin of the latter is traced down and discussed at length, whereas its relation to the all important concept of potentiality in forming a coherent picture of the puzzling entangled interconnections among spatially separated systems is also considered. Finally, certain philosophical consequences of quantum nonseparability concerning the nature of quantum objects, the question of realism in quantum mechanics, and possible limitations in revealing the actual character of physical reality in its entirety are explored.

1   Part and Whole in Classical Mechanics
2   Part and Whole in Quantum Mechanics
3   Strong/Relational Form of Nonseparability
4   Weak/Deconstructional Form of Nonseparability
5   The Relation of Nonseparability to Potentiality
6   Some Philosophical Consequences
6.1   The Context-Dependence of Objects
6.2   Active Scientific Realism
6.3   In Principle Limitation of Knowing the Whole in its Entirety

**1. Part and Whole in Classical Mechanics**

An unambiguous discussion of the part-whole relation in classical physics — namely, the relation between a compound system and its constituent subsystems with respect to the interconnection of their properties — requires the technical conception of the formulation of a compound system on the state space of classical mechanics. We briefly note in this respect that in the case of point-like analytic mechanics, the state of a compound system consisting of N point particles is specified by considering all pairs $\{q_{3N}(t), p_{3N}(t)\}$ of the generalized position and momentum coordinates of the individual particles. Hence, at any temporal moment t, the individual pure state of the compound system consists of the N-tuple $\omega = (\omega_1, \omega_2, ... , \omega_N)$, where $\{\omega_i\} = \{q_i, p_i\}$ are the pure states of its constituent subsystems. It is then clear that in the individual, analytical interpretation of classical mechanics maximal knowledge of the constituent parts of a compound system provides maximal knowledge of the whole system (see, for example, Scheibe, 1973, pp. 53-54). Accordingly, every property the compound system has at time t, if encoded in $\omega$, is determined by $\{\omega_i\}$. For instance, any basic physical quantities (such as mass, momentum, angular momentum, kinetic energy, center of mass motion, etc.) pertaining to the overall system are determined in terms of the corresponding quantities of its parts. They simply constitute direct sums of the relevant quantities of the subsystems. Thus, they are wholly specified by the subsystem states. This is a direct consequence of the application of the conservation laws to the parts and the whole of a classical system and the requirement of additivity of the conserved mechanical quantities. Furthermore, derived quantities, such as the gravitational potential energy, that are not additive in the same simple way as the aforementioned basic quantities, can be explicitly calculated in terms of them. In fact, any derived quantity in classical physics, given its general mathematical expression, can be determined in terms of additive, basic physical quantities, i.e. one-point functions, whose values are well specified at each space-time point.

The foregoing concise analysis delimits the fact, upon which the whole classical physics is founded, that any compound physical system of a classical universe can be conceived of as consisting of separable, distinct parts interacting by means of forces, which are encoded in the Hamiltonian function of the overall system, and that, if the full Hamiltonian is known, maximal knowledge of the values of the physical quantities pertaining to each one of these parts yields an exhaustive knowledge of the whole compound system. In other words, classical physics obeys a *separability principle* that can be expressed schematically as follows:

> *Separability Principle*: The states of any spatio-temporally separated subsystems $S_1$, $S_2$, ..., $S_N$ of a compound system S are individually well defined and the states of the compound system are wholly and completely determined by them and their physical interactions including their spatio-temporal relations (cf. Howard, 1989; Healey, 1991).

Let us now briefly consider the behaviour of the states of a compound system in classical statistical mechanics in relation to the aforementioned separability principle. In the statistical framework of classical mechanics, the states of a system are represented by probability measures $\mu(B)$, namely, non-negative real-valued functions defined on appropriate Borel subsets of phase space. Since the pure state $\omega$ of an individual classical system cannot be known with absolute precision, $\mu(B)$ is simply interpreted as the probability that the individual pure state $\omega$ is more likely to be in the Borel set $B$ of phase space than in others. Consequently, a classical statistical state merely represents an estimate for the pure state of an individual system, where the probability measure $\mu$ describes the uncertainty of our estimation. It can now easily be shown that given the statistical state $\mu$ of a compound system, the states $\{\mu_i\}$, i = 1,2, ..., N , of its subsystems are uniquely determined by $\mu$. In classical theories the set of all statistical states (i.e., probability measures) is a *simplex*, so that every statistical state is the resultant of a unique measure supported by its constituent pure states (e.g., Takesaki, 1979, ch. IV6). In other words, every classical statistical state admits a *unique* decomposition into mutually disjoint pure states, which, in turn, can be interpreted epistemically as referring to our knowledge about an individual system. Thus, every classical statistical state specifies in fact a unique ensemble of pure states in thorough harmony with the separability principle of classical physics.

We finally note with respect to the classical field-theoretical viewpoint, including general relativity, that it does conform to the aforementioned separability principle. The essential characteristic of any field theory, regardless of its physical content and assumed mathematical structure, is that the values of fundamental parameters of a field are well-defined at every point of the underlying manifold (e.g., Einstein, 1971, pp. 170-171). For instance, exhaustive knowledge of the ten independent components of the metric tensor at each point within a given region of the space-time manifold, completely determines the gravitational field in that region. In this sense, the total existence of a field in a given region is contained in its parts, namely, its points. Thus, in attributing physical reality to point-values of basic field parameters, a field theory proceeds by tacitly assuming that a physical state is ascribed to each point of the manifold, and this state determines the local properties of this point-system. Furthermore, the compound state of any set of such point-systems is completely determined by the individual states of its constituents. Consequently, the separability principle is incorporated in the very structure of field theories; in other words, classical field theories necessarily satisfy the separability principle.

## 2. Part and Whole in Quantum Mechanics

In contradistinction to classical physics, standard quantum mechanics systematically violates the conception of separability.[1] The source of its defiance is due to the tensor-product structure of Hilbert-space quantum mechanics and the superposition principle of states. As a means of facilitating the discussion, we shall mostly consider in the sequel the simplest case of a compound system S consisting of a pair of subsystems $S_1$ and $S_2$, since the extension to any finite number is straightforward. In quantum mechanics, by clear generalization of the case of an individual system, to every state of a compound system corresponds a density operator W — a self-adjoint, positive, trace class operator whose trace is equal to 1 — on a tensor-product Hilbert space of appropriate dimensionality. In particular, to every pure state $|\Psi>$ of the compound system corresponds an idempotent density operator $W=W^2$, namely, a projection operator $P_{|\Psi>}= |\Psi><\Psi|$ that projects onto the one-dimensional subspaces $H_{|\Psi>}$ of that product space.

If, then, $W_1$ and $W_2$ are density operators corresponding respectively to the quantal states of a two-component system $S_1$ and $S_2$, the state of the compound system S is represented by the density operator $W = W_1 \otimes W_2$ on the tensor-product space $H_1 \otimes H_2$ of the subsystem Hilbert spaces. The most important structural feature of the tensor-product construction, intimately involved with the nonseparability issue, is that the space $H_1 \otimes H_2$ is not simply restricted to the topological (Cartesian) product of $H_1$ and $H_2$, but includes it as a proper subset. This means that although all vectors of the form $|\psi_i> \otimes |\varphi_j>$ ($\{|\psi_i>\} \in H_1$, $\{|\varphi_j>\} \in H_2$) belong to the tensor-product space, not all vectors of $H_1 \otimes H_2$ are expressible in this form. For, by the principle of superposition there must be linear combinations of vectors $|\psi_i> \otimes |\varphi_j> + |\psi'_i> \otimes |\varphi'_j> + ...$ , which belong to $H_1 \otimes H_2$, but, cannot, in general, factorise into a single product. Put it another way, the metric of the tensor-product space ensures that every vector $|\Psi> \in H_1 \otimes H_2$ can indeed be written as $|\Psi> = \sum_{i,j} c_{ij} (|\psi_i> \otimes |\varphi_j>)$. It does not guarantee, however, that there exist two sets of complex numbers, $\{a_i\}$ and $\{b_j\}$, such that $c_{ij} = a_i b_j$. If this happens, then clearly $|\Psi> = |\psi_i> \otimes \varphi_j>$. In such a case, the state of the compound system is called a *product state*: a state that can always be decomposed into a single tensor-product of an $S_1$-state and an $S_2$-state. Otherwise, it is called a *correlated state*, or in Schrödinger's, 1935a, locution, an *entangled state*.

It is worthy to signify that in quantum mechanics a compound system can be decomposed, in a *unique* manner, into its constituent subsystems if and only if the state of the compound system is of a product form (e.g., von Neumann, 1955). In such a circumstance the constituent subsystems of a compound system behave in an independent and uncorrelated manner, since for any two physical quantities $A_1$ and $A_2$ pertaining to subsystems $S_1$ and $S_2$, respectively, the probability distributions of $A_1$ and of $A_2$ are disconnected: Tr $(W_1 \otimes W_2)(A_1 \otimes A_2)$ = Tr $(W_1 A_1) \cdot$ Tr $(W_2 A_2)$. Thus, correlations among any physical quantities corresponding to the two subsystems are simply non existent. When therefore a compound system is represented by a state of product form, each subsystem possesses an independent mode of existence, namely, a separable and well-defined state, so that the state of the overall system consists of nothing but the sum of the subsystem states. Consequently, the state of the whole is reducible to the states of the parts in consonance with the separability principle of Section 1. This is the only highly particular as well as idealised case in which a separability principle holds in quantum mechanics.

For, even if a compound system at a given temporal instant is appropriately described by a product state, the preservation of its identity under the system's natural time evolution is, from a physical point of view, a considerably rare phenomenon. This is evident by the fact that in order for a compound system to preserve for *all* times a state of product form — $W(t) = W_1(t) \otimes W_2(t)$, for all $t \in \mathbb{R}$ — its Hamiltonian $H$ (i.e., the energy operator of the system) should be decomposed into the direct sum of the subsystem Hamiltonians — $H = H_1 \otimes I_2 + I_1 \otimes H_2$ — and this is precisely the condition of no interaction between $S_1$ and $S_2$. Hence, the slightest interaction among the subsystems of a compound system results in a state for the overall system that is no longer a product of pure vector states, each of which would belong to the Hilbert space of the corresponding subsystem. The time development of the compound system, as determined by the unitary character of Schrödinger's dynamics, still transforms pure states into pure states, but, in general, no longer maps product states into product states. If the physically allowable set of pure states of a compound system is not exhausted by product states, then the smallest interaction of a subsystem with its environment gives rise to an entangled state representation for the whole system. It may appear, in this respect, that the natural realization of the notion of an entangled state is a matter of dynamical interaction, since it presupposes its involvement. As we will argue in Section 5, however, the origin of quantum entanglement is of kinematical rather than dynamical nature. This constitutes a fact of fundamental importance and also the riddle of the problem of nonseparability in quantum mechanics.

## 3. Strong/Relational Form of Nonseparability

**3.1** Let us then consider a compound system S consisting of two subsystems $S_1$ and $S_2$ with corresponding Hilbert spaces $H_1$ and $H_2$. Naturally, subsystems $S_1$ and $S_2$, in forming system S, have interacted by means of forces at some time $t_0$ and suppose that at times $t > t_0$ they are spatially separated. Then, any pure state W of the compound system S can be expressed in the tensor-product Hilbert space in the Schmidt form

$$W = P_{|\Psi>} = |\Psi><\Psi| = \sum_i c_i (|\psi_i>\otimes|\varphi_i>), \qquad ||\ |\Psi>\ ||^2 = \sum_i |c_i|^2 = 1 , \qquad (1)$$

where $\{|\psi_i>\}$ and $\{|\varphi_i>\}$ are orthonormal vector bases in $H_1$ (of $S_1$) and $H_2$ (of $S_2$), respectively. The prominent fact about representation (1) is that it involves only a single summation. It was introduced in this context by Schrödinger, 1935a, in an informal manner in his famous 'cat paradox' paper.

Obviously, if there is just one term in the W-representation of the compound system, i.e., if $|c_i| = 1$, the state $W = |\psi>\otimes|\varphi>$ is a product state and thus represents an individual state of S. If, however, there appear more than one term in W, i.e., if $|c_i| < 1$, then there exist correlations between subsystems $S_1$ and $S_2$. It can be shown in this case that there are no subsystem states $|\xi>$ ($\forall\ |\xi>\in H_1$) and $|\chi>$ ($\forall\ |\chi>\in H_2$) such that W is equivalent to the conjoined attribution of $|\xi>$ to subsystem $S_1$ and $|\chi>$ to subsystem $S_2$, i.e., $W \neq |\xi> \otimes |\chi>$.[2]

Thus, when a compound system such as S is in an entangled state W, namely a superposition of pure states of tensor-product forms, neither subsystem $S_1$ by itself nor subsystem $S_2$ by itself is associated with an individual definite state. The normalised unit vectors $|\psi_i>$, $|\varphi_i>$ belonging to the Hilbert space of either subsystem are not eigenstates of W. If descriptions of physical systems are restricted to the state vector assignment of states, then, strictly speaking, subsystems $S_1$ and $S_2$ have no states at all, even when $S_1$ and $S_2$ are spatially separated. Only the compound system is in a definite pure state W, represented appropriately by a state vector in the tensor-product Hilbert space of S. Maximal knowledge of the whole system, therefore, does not allow the possibility of acquiring maximal knowledge of its component parts, a circumstance with no precedence in classical physics.

Since on the state vector ascription of states, neither subsystem $S_1$ nor subsystem $S_2$ has a state vector in S, it is apparent that the state W of the compound system cannot be reproduced on the basis that neither part has a state. Thus, the separability principle of Section 1 is flagrantly violated. In this respect, the entangled state W establishes a certain kind of quantum nonseparability that we call *strong nonseparability*. On this view an extended compound system is, strictly speaking, an individual quantum system that consists of 'no parts'. The last expression should be understood in the sense that the 'parts' are assigned no definite states. By means of a maxim, strong nonseparability may be described by saying that 'there exists only the 'whole' and not the 'parts'' or 'the 'parts' are devoid of their individuality'. Strong nonseparability casts severe doubts on the existence of isolated (sub)systems and the applicability of the notion of atomism, in the sense that the parts of a quantum whole no longer exist as precisely defined individual entities (in this connection, see also the ending of Section 4.1).

**3.2** Alternatively, when considering an entangled compound system such as S, each subsystem $S_1$ [$S_2$] may be regarded to 'have' a state, but at the expense of being specified only when reference is made to the partner subsystem $S_2$ [$S_1$], via the total information contained in S. Accordingly, each subsystem may be viewed to derive its existence only from its 'behaviour pattern' within the whole. To express this part-whole relation clearly, let us consider an important class of compound quantum systems that form the prototype of EPR-correlated systems, namely, spin-singlet pairs. Let then S be a compound system consisting of a pair ($S_1$, $S_2$) of spin-1/2 particles in the singlet state

$$W = 1/\sqrt{2} \{|\psi_+\rangle \otimes |\varphi_-\rangle - |\psi_-\rangle \otimes |\varphi_+\rangle\},$$

where $\{|\psi_\pm\rangle\}$ and $\{|\varphi_\pm\rangle\}$ are orthonormal bases in the two-dimensional Hilbert spaces $H_1$ and $H_2$ associated with $S_1$ and $S_2$, respectively. In such a case, it is quantum mechanically predicted and experimentally confirmed that the spin components of $S_1$ and of $S_2$ have always opposite spin orientations; they are perfectly anticorrelated. Whenever the spin component of $S_1$ along a given direction is found to be $+1/2$ (correspondingly $-1/2$ ), then the spin component of $S_2$ along the same direction must necessarily be found to be $-1/2$ (correspondingly $+1/2$ ), and conversely. Moreover, due to the rotational invariance of the spin-singlet state, this holds independently of the choice of direction with respect to which the state of spin is expressed.

The spin-singlet pair of particles ($S_1$, $S_2$) clearly establishes nonseparability. Particles $S_1$ and $S_2$, after they have interacted in the past to form the compound system S, remain anticorrelated (maximally entangled) even if they are spatially separated and *no* longer interacting. The state W of the system they formed cannot be reproduced by the states of its parts and their spatio-temporal relationships.[3] Subsystems $S_1$ and $S_2$ are correlated in a way that is encoded in the state W of the overall system. The interrelation, the mutual adaptivity between the parts ($S_1$, $S_2$), cannot be disclosed in an analysis of the parts that takes no account of the entangled connection of the whole. Their interlocking relation does not supervene upon any intrinsic or relational properties of the parts taken separately. It may seem odd to consider nonsupervenient relations holding between nonindividuatable relata. However, the important point to be noticed is that within the context of the foregoing analysis the nature and properties of the parts are determined from their 'role' — the forming pattern of the inseparable web of relations — within the whole. Here, the part-whole relationship appears as complementary: the part is made 'manifest' through the whole, while the whole can only be 'inferred' via the interdependent behaviour of its parts. Thus, in the example under consideration, the property of total spin of the whole in the singlet state does indicate the way in which the parts are related with respect to spin, although neither part possesses a definite numerical value of spin in any direction in distinction from the other one. And it is *only* the property of the total spin of the whole that contains *all* that can be said about the spin properties of the parts, because it is only the whole that contains the correlations among the spin probability distributions pertaining to the parts. The same goes of course for the properties of total momentum and relative distance of the whole with respect to the corresponding local properties of its parts. This part-whole relationship we call, after Teller, 1986, *relational nonseparability* constituting an immediate aspect of the strong kind of quantum nonseparability.[4]

It should be underlined that analogous considerations to the paradigmatic case of the singlet state apply to any case of quantum entanglement. Entanglement need not be of maximal anticorrelation, as in the example of spin-singlet pairs. It does neither have to be confined to states of quantum systems of the same kind. As noted in Section 2, entanglement reaches in principle the states of all quantum systems; it is ubiquitous among them (see also, for instance, Redhead, 1995, pp. 61-62). Consequently, any entangled compound system possess properties which do not reduce to, or supervene on the properties of its component subsystems. As a means of exemplification, within the spin-singlet state W of S is encoded the property that S has total spin zero. From a physical point of view, this derives from the interference (the definite phase interrelations) with which the subsystem states $|\psi\rangle$ and $|\varphi\rangle$ — or, more precisely, the two product states $|\psi_+\rangle \otimes |\varphi_-\rangle$, $|\psi_-\rangle \otimes |\varphi_+\rangle$ — are combined within W. This, in turn, leads not only to the subsystem interdependence of the type described above, but also to conservation of the total angular momentum for the pair $(S_1, S_2)$ of spin-1/2 particles, and thus to the property of definite total spin of value zero for the compound system S.

The latter is a *holistic* property of S: it is not determined by any physical properties of its subsystems $S_1$, $S_2$ considered individually. Specifically, the property of S 'having total spin zero' does not supervene on the spin properties of $S_1$ and of $S_2$, since neither $S_1$ nor $S_2$ has any definite spin in the singlet state. Moreover, the probability distributions concerning spin components of $S_1$ and of $S_2$ along some one direction do not ensure, with probability one, S's possession of this property. Neither the latter could be understood or accounted for by the possibility (that a strict adherent of reductionism may favour) of treating $S_1$ and $S_2$ separately at the expense of postulating a relation between them as to the effect of their spin components 'being perfectly anticorrelated' (see also Healey, 1994). For, while 'having total spin zero' is an intrinsic physical property of the compound system S in the nonseparable state W, the assumed relation is not an intrinsic physical relation that $S_1$ and $S_2$ may have in and of themselves. That is, although the relation of perfect anticorrelation is encoded within state W, ascribing this relation to individual parts of a system is not tantamount to being in state W. The relation of perfect anticorrelation is inherent to the entangled state W itself, whose nonseparable nature reflects furthermore the quantum exclusion principle for half-integer spin particles.

This is precisely the delicate point with entangled correlations in Hilbert-space quantum mechanics: they cannot be explained in terms of preassigned relations or interactions among the parts; their existence cannot be traced back to any interactions. As aforementioned in Section 2, whereas interaction gives rise to entanglement, entanglement itself needs no recourse to interaction for its being established. Interaction is a sufficient but not a necessary condition for entanglement. Quantum entanglement does occur in the absence of any interactions. Due to that the entangled correlations among the states of various physical systems do not acquire the status of a causally dependent relation (cf. Scheibe, 1991, p. 228). Their delineation is rather determined by the entangled quantum state itself which refers directly to the whole system. Thus, we are confronted with a genuine quantum mechanical instance of holism: there exist properties of entangled quantum systems which, in a clearly specifiable sense, characterise the whole system but are neither reducible to nor implied by or causally dependent on the local properties of its parts.

## 4. Weak/Deconstructional Form of Nonseparability

**4.1** In view of the radical form of nonseparability on the state vector ascription of states, it is interesting to inquire whether the assignment of statistical states to component subsystems, represented by non idempotent density operators, restores a notion of separability into quantum theory? The question, contrary to some recently expressed views, is answered strictly in the negative.

Although such a response could be formulated in a quite general manner, the clearest way to see this, for present purposes, is by regarding again as the state W of a compound system the singlet state of a pair of spin-1/2 particles ($S_1$, $S_2$) in the familiar development

$$W = 1/\sqrt{2}\ \{|\psi_+\rangle \otimes |\varphi_-\rangle - |\psi_-\rangle \otimes |\varphi_+\rangle\}. \qquad (2)$$

Observe, in consonance with the considerations of Section 3.1, that neither particle $S_1$ nor particle $S_2$ can be represented in W of Eq. (2) by a state vector. However, each particle may be assigned a state, albeit a reduced state, that is given by the partial trace of the density operator W of the compound system. Recall that the reduced state of each particle arises by 'tracing over' the degrees of freedom associated with the Hilbert space representation of the partner particle. Hence, the following density operators

$$\hat{W}_1 = 1/2\ P_{|\psi_+\rangle} + 1/2\ P_{|\psi_-\rangle} \quad \text{and} \quad \hat{W}_2 = 1/2\ P_{|\varphi_+\rangle} + 1/2\ P_{|\varphi_-\rangle} \qquad (3)$$

represent the reduced ('unpolarized') states of spin-1/2 particles $S_1$ and $S_2$, respectively, in state W.

It is not hard to show, however, that the component states $\hat{W}_1$ and $\hat{W}_2$ of (3) could be identical if derived from a compound state W′ that would correspond to the triplet state

$$W' = 1/\sqrt{2}\ \{|\psi_+\rangle \otimes |\varphi_-\rangle\ +\ |\psi_-\rangle \otimes |\varphi_+\rangle\},$$

or to the following pure states (proviso the sign)

$$W'' = 1/\sqrt{2}\ \{|\psi_+\rangle \otimes |\varphi_+\rangle\ \pm\ |\psi_-\rangle \otimes |\varphi_-\rangle\}$$

that yield, in general, different predictions than W or W′ does for certain spin measurements of both $S_1$ and $S_2$ along a given direction; or they still could be identical if derived from the mixed state

$$W''' = 1/2\ \{|\psi_+\rangle \otimes |\varphi_-\rangle\ +\ |\psi_-\rangle \otimes |\varphi_+\rangle\}.$$

Thus, given the states $\hat{W}_1$ and $\hat{W}_2$ of subsystems $S_1$ and $S_2$, respectively, the compound state could equally well be either W or W′, W″, W‴ or in numerous other variations of them. There exists a many-to-one mapping of the subsystem non-pure states to the state of the whole system.[5] Accordingly, the specification of the compound system remains indefinite. For, the compound state W contains correlations between subsystems $S_1$ and $S_2$ that the reduced states $\hat{W}_1$ and $\hat{W}_2$ do not contain. The sort of correlations that is missing corresponds, from a formal point of view, to the tracing out in the specification, for instance, of $\hat{W}_1$ of what might be known about the state of subsystem $S_2$ and about its connection with subsystem $S_1$. It is evident, therefore, that at the level of description in terms of a reduced state, the information obtained by considering, even simultaneously, the two subsystems does not permit the reconstruction of the pure state of the whole system. In this sense, it may be said that the 'whole' (e.g., system S) is, in a non-trivial way, more than the combination of its 'parts' (e.g., subsystems $S_1$ and $S_2$), this being the case even when these parts occupy remote regions of space however far apart. We shall be referring to this state of affairs by using the term *weak nonseparability*. The usage of this term is justified by the fact that although on the density operator ascription of states each subsystem possess a state independently of the other, still the state of the whole system fails to be exhausted by a specification of the states of the parts and their permissible correlations, bearing in mind that a 'part' or a 'subsystem' within a holistic quantum system acquires no individual independent existence.

To clarify this latter point further, we emphasise that a subsystem state of a compound (entangled) system can be determined only by means of a non-pure state; in fact, a non-pure reduced state. Observe, for instance, in the case under consideration, that the compound state W of Eq. (2) is an entangled pure state; yet each of the subsystems considered individually can only be specified up to the non-pure states of (3). The latter arise as a restriction of the overall pure state of the compound system to the observables pertaining to a component subsystem.[6] Any subsystem state in this situation is exclusively defined at the level of the overall system; there is no individual state for a component subsystem alone. For, there exists no justification in regarding the prescribed reduced states of Eq. (3) as being associated with any specific ensemble of pure states with corresponding eigenvalue probabilities. The reference of a reduced state is only of a statistical, epistemic nature. It simply reflects the statistics derived by a series of measurements performed on a given component subsystem. As a result, the ascription of a reduced state to a subsystem either *presupposes* the suppression or *assumes* negligence of the entangled correlations between subsystems or (sub)systems and their environments. In fact, whenever the pure entangled state of a compound system is decomposed in order to represent subsystems, the effect is a representation in terms of reduced (statistical) states of those subsystems.

**4.2** Acknowledging as a starting point the fact that the nonpurity of a (sub)system's state is due to the breaking of the entangled correlations between the (sub)system concerned and its environment, a certain version of quantum nonseparability arises that may be called, for present purposes, *deconstructional*. It is a conception of nonseparability of the weak kind that transcends the deeply seated presupposition of a whole consisting of given parts which then abide by some form of relation to the whole. In adopting a 'top-down' thinking, the notion of deconstructional nonseparability is based on the fact that the decomposition of a whole into elementary constitutive parts is not, in general, unique.

It is a formal feature of Hilbert-space quantum mechanics that given N appropriate systems $S_1, S_2, ..., S_N$ we can construct uniquely (up to an isomorphism) a compound system $S = \sum_{i=1}^{N} S_i$ on the tensor-product Hilbert space H of the subsystem spaces $H_i$. Given, however, a compound system S, how can we decompose the corresponding Hilbert space H as a tensor-product $H = H_1 \otimes H_2 \otimes ... \otimes H_N$ of the subsystem spaces $H_i$, and such that an observable A of S can be expressed in the canonical form $A = A_1 \otimes A_2 \otimes ... \otimes A_N$ of suitable observables of the subsystems $S_i$? Apparently, in quantum mechanics there are no reasonable criteria that would guarantee the *uniqueness* or even the *existence* of such a tensor-product decomposition of the whole system into its component subsystems (e.g., Primas, 1993; 1994). On the operational level of description, the question of existence presupposes the feasibility of the kinematical independence between a component subsystem of interest and an appropriate measuring system (including its environment), or more generally, it presumes the separation between the observer and the observed, the knowing subject and the object of knowledge.

Recall that from a fundamental viewpoint of quantum theory the physical world appears to us as an unbroken whole. It does not present itself already separated. We have to separate it. In order to be able to obtain any kind of description, to speak about an object, or refer to experimentally accessible facts the underlying wholeness of nature should be decomposed into interacting but disentangled subsystems, namely into measured objects and uncorrelated observers (measuring apparata) with no (or insignificantly so) holistic correlations among them. This subject-object separation is sometimes metaphorically called the Heisenberg cut (e.g., Heisenberg, 1958, p. 116).

The presuppositions of the latter are automatically satisfied in classical physics, in consonance with the separability principle of Section 1. In a nonseparable theory like quantum mechanics, however, the concept of the Heisenberg cut acquires the status of a methodological regulative principle through which access to empirical reality is rendered possible. The innovation of the Heisenberg cut, and the associated separation of a quantum object from its environment, is mandatory for the description of measurements. It is, in fact, necessary for the operational account of any directly observable pattern of empirical reality. The very possibility of devising and repeating a controllable experimental procedure presupposes the existence of such a subject-object separation. Without it the concrete world of material facts and data would be ineligible; it would be conceived in a totally entangled manner.

The feasibility of applying a Heisenberg cut depends critically on a suitable tensor-product decomposition of material reality. In the mathematical framework of quantum mechanics, the latter is not given a priori according to the demands of an Archimedean viewpoint, as reductive atomism would normally require. In the formal codification of the theory there appears neither an intrinsic tensor-product structure, nor a universally valid principle for determining the kind of the decomposition. On the contrary, the appropriate choice of the latter is contingent, in a non trivial manner, upon the context of discourse, namely the idealizations or abstractions that are deliberately made in the scientific process to distinguish between 'relevant' and 'irrelevant' features (see, especially, Section 6.1). As Heisenberg, 1971, p. 58, put it, ''what we observe is not nature itself, but nature exposed to our method of questioning''.[7] In this respect, what we define as a 'part' or as a 'subsystem' within a holistic system is simply a specific pattern (represented quantum mechanically by a non-pure state) that arises by abstracting or cutting through its entangled correlations to the rest of the world, and this may be done in different ways depending on the adopted method of questioning, on the chosen context of investigation.

However, this does not imply that any conceivable contextual decomposition is pertinent. Holistic systems are not in toto vacuous of internal differentiations. The choice of a tensor-product decomposition does satisfy certain physical constraints resulting from the dynamics and the initial state of the compound system. In this sense, certain decompositions of a holistic system into subsystems are more natural, more stable than others.[8] For instance, it appears tenable that the decomposition of the pure state of an electron pair in an EPR-type situation prior to a measurement procedure leads naturally to the reduced states of two single electrons rather than arbitrary other subsystem states or peculiar mixtures of them after the completion of the measurement. In the generic case, however, and essentially for systems whose underlying state space is not locally compact (e.g., systems with infinitely many degrees of freedom), there exists a multitude of physically attainable and physically non-isomorphic decompositions of a whole into (component) parts.

Thus, the nonseparable structure of quantum mechanics suggests that for consistency reasons it is no longer admissible to speak of a quantum whole as being made out of elementary components, but rather that it can be decomposed into or can be described — under suitable circumstances — in terms of such components (Primas, 1998, and references therein). It is the decomposition which 'produces' the possible components of a quantum whole by means of an experimental intervention that suppresses (or sufficiently minimizes) the entangled correlations among them. In this sense, it is justified to say that the process of decomposition provides a level of description to which one can associate a separable concept of reality whose elements are commonly experienced as distinct, well-localised objects.

## 5. The Relation of Nonseparability to Potentiality

In providing a coherent picture of nature that goes beyond the computational rules of current quantum mechanics, it is only natural to investigate as exhaustively as possible into the source or underlying conditions of quantum nonseparability. From a formal point of view, accountable for this counter-intuitive feature is both the tensor-product structure of Hilbert space and the quantum mechanical principle of the superposition of states. In relation to the former, as has already been noted in Section 2, in the tensor-product Hilbert space of a compound system, the states which are products of well-defined states of its constitutive subsystems form a set of measure zero. Thus, strictly speaking, from a quantum theoretical viewpoint there exist no separate systems in the world, apart from the world itself. In relation to the latter, the superposition principle typically states that any number of pure states in Hilbert space can be superposed to generate a new pure state: that is, if $|\psi_1\rangle$, $|\psi_2\rangle$, ... is an arbitrarily large number of unit vectors in H pertaining to a quantum system, then any linear combination

$$|\Psi\rangle = c_1|\psi_1\rangle + c_2|\psi_2\rangle + ... + c_k|\psi_k\rangle + ... \qquad \{c_i\} \in C, \quad \sum_i |c_i|^2 = 1, \qquad (4)$$

constitutes another pure state $|\Psi\rangle$ in H representing also a physically possible state of the system.

To disclose the relation of the superposition principle of states to the question of nonseparability, the states of a system involved in a superposed vector of the form (4) should be explicitly interpreted so as to refer to *potentially realised* (through the measurement process or 'spontaneously' in nature) states $|\psi_i\rangle$, each possessing a probability amplitude $c_i$ of occurrence. The principle of superposition of states is inseparably linked with the *interference* of such probability amplitudes, $c_i c_i^*$, reflecting the nature of the *interrelations* among the states of a quantal system. Accordingly, any physical variable *A* that is associated with a superposed state $|\Psi\rangle$ possess *no* definite value at all, rendering unattainable thereby a Boolean yes-no classification of *A* in $|\Psi\rangle$. In other words, for any physical variable *A* in a superposed state $|\Psi\rangle$ of eigenstates of *A* and any proposition *P* concerning *A*, it is not true that either *P* holds or its complement $I - P$ holds. In such a circumstance, the possible numerical values of *A* are *objectively indeterminate* and not simply unknown. The objectivity of the indeterminacy of *A* stems from the fact that the probabilities of the various possible outcomes or realizations of *A* are designated by the superposed state itself; a feature with no analogue in classical mechanics. True, a notion of superposition is also present for classical mechanic systems, but it is of a radically different nature from any occurring in quantum theory.

In classical mechanics the only kind of superposition possible for a system is a *mixture* of actual, fixed in advance states. No classical mechanic pure state is a superposition of other pure states (e.g., Gudder, 1970). For, there exists no way of combining a number of fine-grained cells, idealised as points in phase space, to produce yet another fine-grained cell and hence a pure state. The classical mechanic states enjoy actual existence. They are real and not potential. In fact, a distinction between the potentially possible and the actually realised is rendered obsolete in classical mechanics by virtue of the absence of an element of chance in the measurement process and the unique (principally causal) determination of the dynamical evolution at each temporal instant. In classical physics all that is potentially possible is also realised in the course of time, independently of any measuring interventions.

It is a distinctive feature of quantum mechanics, on the other hand, that a pure state can be defined independently of measurement only by a probability distribution of potentially possible values which pertain to the physical quantities of a system. In this sense, the quantum state may be construed, regardless of any operational procedures, as representing a network of potentialities, namely, a field of *potentially possible* and not actually existing events. The double modality used here does not simply mean to characterize a transition from potentiality to actuality or from a situation of indefiniteness to definiteness. It also intends to signify that quantum mechanical potentialities condition but do not control the production of actual events.

The concept of quantum mechanical potentiality corresponds to the tendency of a quantum system to display a certain measurement result out of a set of multiplied possibilities in case a suitable measurement is made.[9] When, for instance, in the standard EPR-Bohm example, a pair of particles ($S_1$, $S_2$) is in the spin-singlet state, *no* spin component of either particle exists in a precisely defined form. All three spin components of each particle, however, coexist in a *potential* form and any one component possess the tendency of being actualized at the expense of the indiscriminacy of the others if the associated particle interacts with an appropriate measuring apparatus. As soon as such an interaction takes place, for example, at time $t_o$, and the spin component along a given direction of , say, particle $S_1$ is measured and found equal to $+1/2$ , the operation of state vector reduction, due to the destruction of the superposition bonds between the tensor-product states involved, imparts to particle $S_2$ a potentiality: that of inducing, with probability one, the opposite value $-1/2$, if and when, at a time $t>t_o$, particle $S_2$ is submitted to an appropriate measurement of the same component of spin as $S_1$. Thus, the spin-singlet state, furnishing the standard EPR-Bohm example, describes the entanglement, the inseparable correlation of potentialities.

The singlet state (as any entangled state) represents in essence a set of potentialities whose content is not exhausted by a catalogue of actual, preexisting values that may be assigned to the spin properties of $S_1$ and of $S_2$, separately. It may be worthy to observe in this connection that in the EPR-Bohm example no spin property of particle $S_2$ enjoyed a well-defined value prior to the measurement performed on $S_1$ and a different value in the sequel. The only change occurring in $S_2$ concerns the transition of a spin property from a potentially possible value to an actually realised value. If the actualized values of all spin properties of $S_2$ were predetermined, then the behaviour of $S_2$ would be fixed in advance regardless of any reference to system $S_1$, and vice versa. Consequently the behaviour of the compound spin-singlet pair would be reducible to the behaviour of its parts. A sense of separability would also had been demonstrated if the actualized values of both $S_1$ and $S_2$ were independent or owed their existence to a common cause in the past (e.g., Butterfield, 1989). Thus, quantum mechanical nonseparability is subject to both the actualization of potentialities and the confinement of actualization to the dictates of correlation.

The concept of quantum mechanical potentiality should not be classified under an epistemic category of apprehension. It does not refer to someone's deficient knowledge as to the exact nature of a given object, but it belongs to the mode of existence of the object itself. It characterizes the degree of realisation of a potentially possible event determined by objective physical conditions, namely, by the internal properties of the object and the specified experimental conditions. Quantum mechanical potentialities are physically real and objective not only in the sense of designating the disposition of an object to expose certain properties when submitted to given conditions, but also in the sense of interfering, under certain circumstances as in quantum coherence or quantum entanglement, with one another (cf. Heisenberg, 1958, p. 185; Popper, 1980, p. 309).

Thus, when confronting a compound system in an entangled state, one may conceive that the potentialities of subsystems $S_1$ and $S_2$, consisting the whole system, interfere with each other in such a way that the probability of a certain result of a measurement performed on $S_1$ is dependent upon the result of a measurement performed on $S_2$, and conversely.[10] Or even more acutely, as exemplified in the physically important case of spin anticorrelation, the actualization of an arbitrary spin component of $S_1$ entails the actualization of the corresponding spin component of $S_2$ with precisely defined value. Numerous experimental results, whose quantitative character is based on a Bell-type inequality, strongly testify that this aspect of interfering potentialities does take part in nature, even when $S_1$ and $S_2$ are spatially separated (e.g., Aspect et al., 1982; Tittel et al., 1998). Furthermore, as has been repeatedly shown, there seems to be no way of utilizing the quantum mechanical transition from potentiality to actuality for the purpose of establishing a superluminal communication procedure between $S_1$ and $S_2$ that would violate special relativity.[11]

It remains an open problem, however, to provide a detailed description of how this transition from possible to actual occurs in nature. The problem of the actualization of quantum mechanical potentialities, known also as the problem of measurement, forms the fundamental difficulty that pervades all quantum theory since its inception. Despite of the intense effort that has been devoted to this problem and the achievement of several significant partial results, there seems to be no straightforward route towards a satisfactory, fully coherent, solution to the problem (e.g., Karakostas, 1994; Karakostas and Dickson, 1995).

## 6. Some Philosophical Consequences
### 6.1 The context-dependence of objects

Regardless of the multifaceted difficulties encountered within a realistic framework of a quantum theory of measurement, and the associated want of an explanation of the potentiality to actuality conversion, quantum mechanics definitely shows that nonseparability constitutes a basic feature of the quantum world. It is worthy to signify that from a foundational viewpoint, quantum mechanical nonseparability and the sense of quantum holism arising out of it refer to a context-independent, or in D' Espagnat's, 1995, scheme, observer- or mind-independent reality. The latter is operationally inaccessible. It pertains to the domain of entangled correlations, potentialities and quantum superpositions obeying a non-Boolean logical structure. Here the notion of an object, whose aspects may result in intersubjective agreement, enjoys no a priori meaning independently of the phenomenon into which is embedded. As already noted in Section 4.2, well-defined separate objects (and their environments) are generated by means of a Heisenberg cut, namely through the process of a deliberate projection of the holistic non-Boolean domain into a Boolean context that necessitates the suppression (or minimization) of entangled correlations between the object-to-be and the environment-to-be (e.g., a measuring apparatus).

The Heisenberg cut is inevitably contextual. It aims at establishing a regulative link between the conditions for an objective description of experience and the inherently nonseparable structure of quantum mechanics. Once Heisenberg's cut is drawn, the kinematic independence of the measuring apparatus is a necessary condition for the knowing subject to record objective information about the object. For, if the state-properties of the apparatus were entangled with the object under measurement, then, the resulting information would be relative. The very definition of physical terms would suffer from an indefinite regress of measurement-context dependency. In such a circumstance, we would not be able to make sense of our experience, to interpret in an exact manner the measurement results, or to interpret the formalism predicting those results — and the theoretical description would have no consequences. A precise formulation of concepts requires the measuring apparatus/knowing subject to be clearly separated from the content to be represented. This content consists of objects. The subject cannot be regarded as one more object, admissible to further analysis. It is, rather, the knowing subject that specifies other things as objects of knowledge. There must be therefore a clear, although artificial, separation between them so as to allow for intrinsic acts of abstraction, of disassociation, of stabilization, and finally, of registration. In this sense, a physical system may account as a measuring device (an extension of the knowing subject) only if it is not holistically correlated or entangled with the object under measurement.

Consequently, any atomic fact or event that 'happens' is raised at the observational level only in conjunction with the specification of an experimental arrangement[12] — an experimental context that conforms to a Boolean domain of discourse — namely to a set of observables co-measurable by that context. In other words, there cannot be well-defined events in quantum mechanics unless a specific set of co-measurable observables has been singled out for the system-experimental context whole (e.g., Landsman, 1995). Or to put it differently, due to the underlying quantum nonseparability the world appears as a complex whole. However, once a particular question is put to nature and therefore a given context is specified the oneness of the whole falls apart into apparent parts. Thus, whereas quantum wholeness refers to an *inner-level* of reality, a mind-independent reality that is operationally elusive, the introduction of a context is related to the *outer-level* of reality, the empirical reality that results as an abstraction in the human perception (Karakostas 2003). The experimental context offers precisely the conditions on the basis of which a quantum event manifests itself. I.e., the experimental context operates as a formative factor for the production of an event.

Consequent upon that the nature of quantum objects is a *context-dependent* issue with the experimental procedure supplying the physical context, the necessary frame for their manifested appearance. This seems to be reminiscent of Kant's view that the concept of an object is a condition of the possibility of its being experienced. However this may be, the meaning of the term reality in the quantum realm cannot be considered to be determined by what physical objects really are. In fundamental quantum mechanics, there are no objects in an intrinsic, absolute sense. Quantum mechanics describes material reality in a substantially context-dependent manner. The classical idealisation of sharply individuated objects possessing intrinsic properties and having an independent reality of their own breaks down in the quantum domain. Quantum attributes can reasonably be regarded as pertaining to an object of our interest only within a framework involving the experimental conditions. The latter provide the necessary conditions whereby we make meaningful statements that the properties we attribute to quantum objects are part of physical reality. Without the consideration of an appropriate context the concept of a quantum object remains unspecified. Or to put the same in different words, any determination of an object in microphysics, as far as state-dependent properties are concerned, is meaningful only within a given context. I.e., objects are *context-dependent* entities.

Accordingly quantum objects can not be conceived of as 'things-in-themselves', as 'absolute' building blocks of reality. Instead, they represent carriers of patterns or properties which arise in interaction with their experimental context/environment, or more generally, with the rest of the world; the nature of their existence depends on the context into which they are embedded and on the abstractions we are forced to make in any scientific discussion. As already noted, the perceptible separability and localizability of the contextual objects of empirical reality are generated by breaking/abstracting the factually existing entangled correlations of the object concerned with its environment. In this sense, contextual quantum objects may be viewed as 'constructed', 'phenomenal' entities brought forward by the theory. According to considerations of Section 4, they lack individuality and permit only an unsharp description in terms of reduced (statistical) states. They are not, however, inventions of the human mind, nor do they constitute a kind of 'noumenal' entities in the Kantian terminology.[13] They reflect objective structural aspects of physical reality with respect to a certain *preselected* context. 'Noumenal' entities, on the other hand, are relevant only in a situation without the application of a Heisenberg cut, without the involvement of any subject-object separation, or any observational interaction. Since the existence of the latter generates contextual objects ('phenomenal' entities), 'noumena' are empirically inaccessible, artificial limiting cases of 'phenomena'. A noumenal entity can produce no informational content that may be subjected to experimental testing without itself being transformed to a phenomenal entity.

## 6.2 Active Scientific Realism

The context-dependency of quantum objects does not contradict a realistic view of the world as usually assumed. Scientific realism and quantum nonseparability are not incompatible, but the relationship between them points to the abandonment of the classical conception of physical reality and its traditional metaphysical presuppositions, most notably, atomism and strict ontological reductionism. Due to the substantially context-dependent description of reality suggested by quantum mechanics, one ought to acknowledge that the patterns of nature emerge and become intelligible through an active participation of the knowing subject that, by itself, is not and can not be perspective-independent. In this sense, the conceptualization of the nature of reality, as arising out of our most basic physical theory, calls for a kind of contextual realism that we call *active scientific realism*.

Active, since it intends to indicate the contribution of thought to experience; it purposes to acknowledge the *participatory* role of the knowing subject in providing the context of discourse; as previously argued, the identification of a specific pattern as an 'object' depends on the process of knowledge, the preselection of an experimental context. And realism, since it maintains that *given* a context, concrete objects (structures of reality) have well-defined properties independently of our knowledge of them. According to the conception of active scientific realism, the experienced or empirical reality is a functional category. Functional to the engaging role of the knowing subject, so that the outer reality is perceived as not something given a-priori, a 'ready-made' objective truth dictated allegedly by an external point of reference, but as something affected by the subject's action. The nonseparable structure of quantum mechanics, the experimentally well-confirmed holistic features arising out of it, and the subsequent context-dependent description of physical reality conflict with the rather comfortable view that the world's contents enjoy an intrinsic and absolute meaning of existence, independent of any involvement of the knowing subject, only to be 'discovered' progressively by him.

The knowing subject — i.e., the experimenter/observer — is free to choose the aspect of nature he is going to probe, the question he selects to investigate, the experiment he is about to perform. The role of the experimenter's choice is inimitable within the framework of quantum mechanics. For, firstly, a choice on the part of the experimenter is not controlled or governed by the laws of contemporary physics, namely quantum theory, secondly, the statistical algorithm of the theory can only be applied if a certain experimental context has first been selected, and, more importantly, thirdly, this element of freedom of choice in selecting the experimenter a context of discourse leads in quantum mechanics to a gradually unfolding picture of reality that is not in general fixed by the *prior* part of the physical world alone. According to the physico-mathematical structure of standard quantum mechanics, the mere freedom to select the experimental context, to choose the question asked to nature and specify the timing that is asked, allows the experimenter to exert an indelible influence on a quantum system. In relation to considerations of Section 5, for instance, such a denotable influence is exerted on the quantum mechanical potentialities for the various possible outcomes obtained by measurements on the system. Indeed, given an initial state vector of a quantum system and the choice of a certain experimental context, the act of measurement injects into the former an increment of information — a particular measurement outcome — associated with the specifications of the latter. The determination of the context characterizes the state vector by a particular subset of potentialities. As soon as the measurement process is accomplished, only one of the potentialities is actualised at the expense of all the others, and the initial state vector qualitatively changes corresponding to a new situation with new possibilities. Thus, in a consequence of measurements, the state vector carries into the future in an *indelible non-reversible* manner the informational content of the choice of each experimental context and it does so in a way that controls statistically the potentialities for the occurrence of subsequent outcomes (Karakostas 2004).

The conception of active scientific realism put forward here aims at precisely capturing the sense in which the knowing/intentional subject is an *effective participant* in the physical world. It aims at appropriately expressing within physics the sense in which this is a "participatory universe", just to recall Wheeler's, 1983, p. 202, phrase. In quantum mechanics, beyond the selection of an experimental context as an act of cognition on the part of the experimenter, the experimenter's volitional decision on what to measure bears also an indelible consequence in bringing about the elementary phenomenon that appears to be happening. By way of a generalized example, if A, B and C represent observables of the same quantum system, so that the corresponding operator A commutes with operators B and C ($[A, B] = 0 = [A, C]$), not however the operators B and C with each other ($[B, C] \neq 0$), then the result of a measurement of A *depends* on whether the experimenter had previously subjected the system in a measurement of the observable B or in a measurement of the observable C or in none of them. This may simply be viewed as a lemma of the Kochen and Specker theorem, 1967, whereas analogous, still more vivid, statements to this consequence can be made through recent investigations of established effects in quantum theory as the quantum Zeno effect and the so-called delayed choice experiments (e.g., Stapp, 2001; Tegmark, 2000 and references therein).

Thus, quantum mechanics undeniably reveals a sense in which the knowing subject participates in bringing 'reality' into being. The conception of active scientific realism — as an interpretative, regulative hypothesis to physical science — attempts to incorporate the human factor into our understanding of reality within which human knowledge forms an indispensable part. It acknowledges the subject-object inherent engagement by integrating the subjective human increments into an objectively existing physical reality. There is an ever growing recognition that the Cartesian ideal of an absolute dichotomy between the human mind and the external world, allowing no intermediate medium, is fundamentally flawed (e.g., Smith, 1996). It leads to an alleged construction of decontaminated by human qualities knowledge — a supposedly perspective-free view of the world — that is deficient to really understand reality. In contradistinction, active scientific realism attempts to convey that the actual relation of the knowing subject to a world is neither of absolute dichotomy, nor of independence or of externality, but of active participancy. Thus in coming to know the physical world, we come to know also ourselves as knowers and consequently understand how our perspective affects or contributes to our conceptualization about the world. And conversely, by knowing how we contribute to the knowledge claims about the world, we more securely identify the informational content the world itself contributes. In this view, human subjectivity and scientific objectivity, within the respective limits of their appropriateness, are no longer contraries; they are jointly achieved.[14]

**6.3 In principle limitation of knowing the whole in its entirety**

Although the relation of the knowing subject to a world is reasonably conceived, as we maintain, as a relation of active participancy, a distinction between subject and object (as phenomena on the ground) should be drawn within the world. As already noted in Section 6.1, for a scientific description of reality that is amenable to experimental tests a subject/object distinction is inevitable. Such an induced subject/object separation represents a necessity in any experimental science including physics. It also constitutes, more specifically, the condition upon which any empirically accessible aspect of entangled quantum correlations is based. It is both interesting and ironic that all empirical evidence concerning quantum nonseparability is obtained by deliberately 'eliminating' that same nonseparability between the object under study and its instrument of observation. In this sense, quantum nonseparability cannot be regarded as an 'observed phenomenon' per se. It can only indirectly be demonstrated or inferred in a separable way, conceptually employing counterfactual reasoning (cf. Atmanspacher, 1994). When the latter is used in Bell-type argumentations, as for instance in the analysis of relevant experimental tests of Bell's inequalities, contradictions may occur with the statistical predictive rules of quantum theory leading to empirically verifiable situations that imply, in turn, the existence of a non-divided, undissectable whole. The formal theory, however, that provides the framework for the systematization of the empirical results is not capable of revealing or describing the actual character of this whole. The latter is strictly speaking indescribable in the sense that any detailed description necessarily results in irretrievable loss of information by dissecting the otherwise undissectable.

Thus, although nonseparability as an inherent element of quantum theory clearly implies that wholeness is a fundamental aspect of the quantum world, nonetheless, neither theory nor experiment can disclose the exact fabric of this wholeness. In this respect, it can safely be asserted that reality thought of as a whole is not scientifically completely knowable, or, at best, it is veiled (cf. D' Espagnat, 1998). It is, in fact, not susceptible to immediate scientific practice; it cannot be the subject-matter of detailed scientific investigation; for, it can neither be quantified nor computed or measured from the outside. Any discussion concerning the nature of this *undissectable whole* is necessarily of an ontological, metaphysical kind, the only confirmatory element about it being the network of interrelations which connect its events. Consequently, we are led to a novel limit on the ability of scientific cognizance as to the actual character of physical reality in its entirety.

The common philosophical assumption concerning the feasibility of a panoptical, Archimedean point of view is rendered illusory in quantum mechanics. In contrast to an immutable and universal 'view from nowhere' of the classical paradigm, quantum mechanics acknowledges in an essential way a perspectival character of knowledge. Quantum mechanics reveals that the hunt of a universal frame of reference for describing physical reality is in vain. Quantum-mechanical reality is nonseparable and her distinctiveness into facts a matter of context. As argued in Section 6.1, access to the non-Boolean quantum world is only gained by adopting a particular Boolean perspective, by specifying a certain Boolean context which breaks the wholeness of nature. Consequently, the description and communication of results of experiments in relation to the non-Boolean quantum world presuppose the impossibility of a perspective-independent account, since one must at the outset single out an experimental context (determined by a set of co-measurable observables for the context-cum-quantum system whole) and in terms of which the definite result of a measurement can be realized.

Be sure there is only one reality, but every description of it presupposes the adoption of a particular point of view. There is no such a thing as a 'from nowhere' perspective or a universal viewpoint. A complete knowledge of the world as a whole would have to provide an explanation of the perception process or the pattern recognition mechanisms of the knowing subject, since this is part of the world. It would have to include within a hypothetically posited ultimate theory an explanation of the conditions for observation, description and communication which we ourselves, as cognizant subjects, are already subjected to. We cannot transcend them. This would be like attempting to produce a map of the globe which contained itself as an element. The usage of the metaphor is meant to convey the conceptually deep fact — reminiscent of Gödel's famous undecidability theorem for axiom-systems in mathematics — that a logically consistent theory can not generally describe its universe as its own object. In particular, the scientific language of our alleged universal ultimate theory would have to be semantically closed, and hence engender antinomies or paradoxes especially in relation to self-referential descriptions, as in the case of von Neumann's account of quantum measurement that leads to an infinite regress of observing observers (e.g., Chiara, 1977; Peres and Zurek, 1982). Be that as it may, the assumption of a 'view from nowhere' appeared realizable prior to quantum mechanics, because in classical physics the validity of separability and unrestricted causality led to the purely reductionist presumption that one could consistently analyze a compound system into parts and consequently deduce the nature of the whole from its parts. Since the part could be treated as a closed system separate from the whole, the whole could ultimately be described — by applying the conservation laws of energy, momentum and angular momentum — as the sum of its parts and their physical interactions, and hence the knowing subject would achieve its knowledge of physical reality from the 'outside' of physical systems.

In the quantum theoretical framework that picture is no longer valid. As we have extensively argued, the consideration of physical reality as a whole — supposedly that a sense is ascribed to this word — can not be comprehended as the sum of its parts in conjunction with the spatiotemporal relations or physical interactions among the parts, since the quantum whole provides the framework for the existence of the parts. Due to the genuinely nonseparable structure of quantum mechanics, whole and parts mutually influence and specify each other; the parts are adjusted by the whole, whereas, in turn, the whole depends on the interconnectedness of its parts. Hence, the legendary notion of the classical paradigm that wholes are completely describable or explicable in terms of their parts is no longer defensible. In classical mechanics, given any compound physical system, the dynamics of the whole was regarded as being reducible to the properties of the parts. In quantum mechanics the situation is actually reversed; the identity and properties of the parts can ultimately be accounted only in terms of the dynamics of the whole. In a truly nonseparable physical system, as in an entangled quantum system, the part does acquire a different identification within the whole from what it does outside the whole, in its own 'isolated', separate state (see esp. Secs. 3.2 & 4.1). Thus, for instance, in the strong/relational view of nonseparability no isolated spin-1/2 particle, e.g. a 'free' or 'bare' electron, can be identified with the spin state of either member of a pair of electrons in the singlet state, whereas, in the weak/deconstructional view of nonseparability no state preparation procedure could have produced such an isolated particle in a reduced state whose status against the singlet state is necessarily only of an epistemic nature. So, in either case, the whole state cannot be regarded as being made up of individual, isolated, separate particle states.

When all is said and done, the present situation in physics suggests that the natural scientist as a conscious being may operate within a mild form of the reductionist paradigm in trying to analyse complex objects in terms of parts with the absolute certainty, however, that during the process the nature of the whole will not be disclosed. The value of the reductionistic concept as a working hypothesis or as a methodological tool of analysis and research is not jeopardized at this point,[15] but ontologically it can no longer be regarded as a true code of the actual character of the physical world and its contents. Quantum mechanical nonseparability strongly suggests that the functioning of the physical world cannot just be reduced to that of its constituents thought of as a collection of interacting but separately existing localized objects. Any coherent conceptualization of the physical world that is compatible with the predictions of quantum mechanics requires us to view the world, in the expression of Heisenberg, 1958, p. 96, "as a complicated tissue of events, in which connections of different kinds alternate or overlay or combine and thereby determine the texture of the whole". Although the latter can hardly be fully knowable, an enlightenment of its actual character may be given by the penetrating power of the theory itself and its future development. In this respect, it is rather safe to conjecture that the conception of quantum nonseparability will be an integral part of the next conceptual revolution in physics and may even be used as a regulative constructive hypothesis guiding the search for the fundamental physics of the future.

# NOTES

[1] In this work we shall make no reference to alternative interpretations of ordinary quantum mechanics as, for instance, Bohm's ontological or causal interpretation.

[2] In a paper related to the Einstein-Podolsky-Rosen argument, Schrödinger remarked with respect to this distinctive feature of nonfactorizability as follows: ''When two systems, of which we know the states by their respective representations, enter into temporary physical interaction due to known forces between them, and then after a time of mutual influence the systems separate again, then they can no longer be described in the same way as before, viz. by endowing each of them with a representative of its own. ... I would not call that one but rather *the* characteristic trait of quantum mechanics, the one that enforces its entire departure from classical lines of thought'' (Schrödinger, 1935b, §1 ).

[3] It is well-known that spin-singlet correlations violate Bell's, 1964, inequalities. We may note in this connection the interesting result of Gisin, 1991, Popescu and Rohrlich, 1992, that for any nonseparable state of a two-component system there is a proper choice of pairs of observables whose correlations do violate Bell's inequality.

[4] The significance of correlations as compared to that of correlata is particularly emphasised by Rovelli, 1996, and Mermin, 1998.

[5] Hughston et al., 1993, provide a constructive classification of all discrete ensembles of pure quantum states that correspond to a fixed density operator.

[6] A non-pure state is usually addressed under the term of a mixed state, mainly because both 'reduced' and 'mixed' states correspond mathematically to a non-idempotent density operator. However, the use of the general term of a mixed state as a substitute of the non-purity of a state is conceptually misleading. A mixed state, in opposition to a reduced state, arises as a particular mixture of different pure states in the (sub)system of interest and thus censures the significance of entangled correlations.

[7] Similar statements can also be found in the writings of Pauli, 1994, pp. 41, 133.

[8] A general explanation of the robustness of certain decompositions of a system as a whole, in terms of stability considerations that could also be taken into account formally, is still lacking. An initiatory attempt towards this direction has been recently indicated by Amann and Atmanspacher, 1998.

[9] The view that the quantum state vector refers to 'possibilities' or 'tendencies', as a certain extension of the Aristotelian concept of 'potentia', has been advocated by Heisenberg, 1958, pp. 42, 53, in his later writings on the interpretation of quantum mechanics, and especially by Fock, 1957, p. 646. Margenau, 1950, pp. 335-337, 452-454, too has used the concept of 'latency' to characterise the indefinite quantities of a quantum mechanical state that take on specified values when an act of measurement forces them out of inditermination. Analogous is Popper's, 1980, ch.9; 1990, ch.1, understanding of attributing properties to quantum systems in terms of objective 'propensities'. Today one of the most eloquent defenders of the appropriateness of the concept of potentiality in interpreting quantum mechanics is Shimony, e.g. 1993, Vol. 2, ch. 11.

[10] The probabilistic dependence of measurement outcomes between spatially separated systems forming a quantum whole corresponds, as an expression of the violation of the separability

principle, to the violation of what has been coined in the Bell-literature as Jarrett's, 1984, 'completeness condition', or equivalently, Shimony's, 1986, 'outcome independence' condition. A detailed description of these conditions would fall outside the scope of the present work. A review of them may be found in Shimony, 1990, or more recently in Howard, 1997.

[11] A recent generalised version of the so-called no-signalling theorem is given by Scherer and Busch, 1993.

[12] It should be pointed out that Bohr already on the basis of his complementarity principle introduced the concept of a 'quantum phenomenon' "to refer exclusively to observations obtained under specified circumstances, including an account of the whole experiment" (Bohr, 1963, p. 73).

[13] For an elaboration on the distinction between 'noumenal' and 'phenomenal' entities in Kant's philosophical scheme, see Allison, 1983.

[14] One can hardly resist in recalling at this point Hermann Weyl's consideration with respect to the subjective-objective relation. Although Weyl's reference is within the context of relativity theory, his remarks are equally pertinent to quantum mechanics as well as to the general relation between the experiencing subject and the external world. He expressly writes: "The immediate experience is subjective and absolute. However hazy it may be, it is given in its very haziness thus and not otherwise. The objective world, on the other hand, with which we reckon continually in our daily lives and which the natural sciences attempt to crystallize by methods representing the consistent development of these criteria by which we experience reality in our everyday attitude — this objective world is of necessity relative; it can be represented by definite things (numbers or other symbols) only after a system of coordinates [the rest-frame of an observer] has been arbitrarily carried into the world. It seems to me that this pair of opposites, subjective-absolute and objective-relative, contains one of the most fundamental insights which can be gleaned from science. Whoever desires the absolute must take the subjectivity and egocentricity into the bargain; whoever feels drawn towards the objective faces the problem of relativity (Weyl, 1949, p. 116).

[15] Quantum holism, as a consequence of nonseparability, should not be regarded as the opposite contrary of methodological reductionism. Holism is not an injunction to block distinctions. A successful holistic research programme has to account apart from nonseparability, wholeness and unity also for part-whole differentiation, particularity and diversity. In this respect, holism and methodological reductionism appear as complementary viewpoints, none can replace the other, both are necessary, none of them is sufficient.